\documentclass[twocolumn,showpacs,aps,prl,superscriptaddress,USletter]{revtex4}
\usepackage{graphicx}
\usepackage{dcolumn}
\usepackage{amsmath}
\usepackage{epsfig}

\input pubboard/babarsym.tex

\newcommand{\romegaomega}{\ensuremath{1.2 \pm 0.3 ^{+0.3}_{-0.2}}}
\newcommand{\romegaphi}{\ensuremath{0.0^{+0.3}_{-0.2} \pm 0.1}}
\newcommand{\ulomegaomega}{\ensuremath{1.9}}
\newcommand{\ulomegaphi}{\ensuremath{0.7}}

\newcommand{\BABARPubYear}    {13}
\newcommand{\BABARPubNumber}  {018}
\newcommand{\SLACPubNumber} {15846}

\newcommand{\calP}{\ensuremath{{\cal P}}}

\newcommand{\pvec}{{\bf p}}

\newcommand{\calB}{\ensuremath{{\cal B}}}

\newcommand{\timesix}{\ensuremath{\times10^{6}}}

\newcommand{\DE}{\ensuremath{\Delta E}}

\newcommand{\xf}{\ensuremath{{\cal F}}}

\newcommand{\thetaT}{\ensuremath{\theta_{\rm T}}}
\newcommand{\costhr}{\ensuremath{\cos\thetaT}}

\newcommand\etal{{\it et al.}}
\newcommand{\half}{\ensuremath{{1\over2}}}

\newcommand{\bfig}{\begin{figure}[htbpc!]}
\newcommand{\efig}{\end{figure}}
\newcommand\bef{\begin{figure}}
\newcommand\edf{\end{figure}}
\newcommand\dbline{\noalign{\vskip 0.10truecm\hrule}\noalign{\vskip 2pt}\noalign{\hrule\vskip 0.10truecm}}
\providecommand{\tbline}{\noalign{\vskip 0.05truecm\hrule\vskip0.05truecm}}

\newcommand\beq{\begin{equation}}
\newcommand\eeq{\end{equation}}
\newcommand\bear{\begin{array}}
\newcommand\enar{\end{array}}
\newcommand\beqa{\begin{eqnarray}}
\newcommand\eeqa{\end{eqnarray}}
\newcommand\ben{\begin{enumerate}}
\newcommand\een{\end{enumerate}}

\newcommand{\omtoppp}{\ensuremath{{\omega\ra\pip\pim\piz}}}

   \newcommand{\rhop}{\ensuremath{\rho^+}}

\newcommand{\UfourS}{\ensuremath{\Upsilon(4S)}}

\newcommand{\fomegaomega}{\ensuremath{\omega\omega}\xspace}
\newcommand{\omegaomega}{\ensuremath{\Bz\ra\fomegaomega}\xspace}
\newcommand{\Bomegaomega}{\ensuremath{\calB(\omegaomega)}\xspace}
\newcommand{\fomegaphi}{\ensuremath{\omega\phi}\xspace}
\newcommand{\omegaphi}{\ensuremath{\Bz\ra\fomegaphi}\xspace}
\newcommand{\Bomegaphi}{\ensuremath{\calB(\omegaphi)}\xspace}

\def\sss{\scriptscriptstyle}
\def\barpd{{\raise.35ex\hbox{${\sss (}$}}--{\raise.35ex\hbox{${\sss )}$}}}
\def\BorBbar{\hbox{$B$\kern-0.85em\raise1.5ex\hbox{\barpd}\hspace{-0.4mm}$^0$}}

\def\figurebox#1#2#3{%
    \def\arg{#3}%
    \ifx\arg\empty
    {\hfill\vbox{\hsize#2\hrule\hbox to #2{\vrule\hfill\vbox to #1{\hsize#2\vfill}\vrule}\hrule}\hfill}%
    \else
    {\hfill\epsfbox{#3}\hfill}%
    \fi}

\begin{document}

\preprint{\babar-PUB-\BABARPubYear/\BABARPubNumber}
\preprint{SLAC-PUB-\SLACPubNumber}

\begin{flushleft}
\babar-PUB-\BABARPubYear/\BABARPubNumber\\
SLAC-PUB-\SLACPubNumber
\end{flushleft}

\title{
{\large \boldmath \bf  Evidence for the decay $\Bz \to \omega\omega$ and search for $\Bz \to \omega\phi$}
}

%
\author{J.~P.~Lees}
\author{V.~Poireau}
\author{V.~Tisserand}
\affiliation{Laboratoire d'Annecy-le-Vieux de Physique des Particules (LAPP), Universit\'e de Savoie, CNRS/IN2P3,  F-74941 Annecy-Le-Vieux, France}
\author{E.~Grauges}
\affiliation{Universitat de Barcelona, Facultat de Fisica, Departament ECM, E-08028 Barcelona, Spain }
\author{A.~Palano$^{ab}$ }
\affiliation{INFN Sezione di Bari$^{a}$; Dipartimento di Fisica, Universit\`a di Bari$^{b}$, I-70126 Bari, Italy }
\author{G.~Eigen}
\author{B.~Stugu}
\affiliation{University of Bergen, Institute of Physics, N-5007 Bergen, Norway }
\author{D.~N.~Brown}
\author{L.~T.~Kerth}
\author{Yu.~G.~Kolomensky}
\author{M.~J.~Lee}
\author{G.~Lynch}
\affiliation{Lawrence Berkeley National Laboratory and University of California, Berkeley, California 94720, USA }
% ...
\collaboration{The \babar\ Collaboration}
\noaffiliation


\begin{abstract}
We describe searches for $B$ meson decays to the charmless vector-vector
final states $\omega\omega$ and $\omega\phi$ 
with $471 \times 10^6 \: B{\kern 0.18em\overline{\kern -0.18em B}}$ pairs produced 
in $e^+ e^-$ annihilation at $\sqrt{s}=10.58$ GeV using the {\mbox{\slshape B\kern-0.1em{\smaller A}\kern-0.1em B{\smaller A\kern-0.2em R}}}
detector at the PEP-II collider at the SLAC National Accelerator Laboratory.
We measure the branching fraction $\mathcal{B}(B^0 \to \omega\omega) = (1.2 \pm 0.3^{+0.3}_{-0.2}) \times 10^{-6}$,
where the first uncertainty is statistical and the second is systematic, corresponding to a significance of 4.4 standard deviations.
We also determine the upper limit $\mathcal{B}(B^0 \to \omega\phi) < 0.7 \times 10^{-6}$ at 90\% confidence level.
These measurements provide the first evidence for the decay $B^0 \to \omega\omega$, and an improvement of the upper limit for the decay
$B^0 \to \omega\phi$.
\end{abstract}


\maketitle

Charmless decays of $B$ mesons to two vector mesons have been of
significant recent interest, in part because of the unexpectedly small
value of the longitudinal polarization component observed in
$B \to \phi\Kstar$ decays~\cite{BabarPhiKstar,BellePhiKstar}.
The resulting large transverse spin component could be due either to
unanticipated large Standard Model (SM) contributions~\cite{VVBSMrefs} or
to non-SM effects~\cite{nSMetc}.  Further information and SM constraints on
these decays can be obtained from measurements of, or limits on, the
branching fractions of related decays, such as \omegaomega and \omegaphi~\cite{oh}.
These latter decays are also important because they contain relatively
unstudied $b \to d$ quark transitions (\omegaomega however is expected to be dominated by
$b \to u$ transitions) and are sensitive to the phase angles $\alpha$ and
$\gamma$ of the Cabibbo-Kobayashi-Maskawa quark mixing matrix~\cite{VVphaserefs}.
Deviations of the observed branching fractions from their SM expectations could
provide evidence for physics beyond the SM.

Theoretical predictions for the SM branching fractions lie in the range (0.5 -- $3) \times 10^{-6}$ for
\omegaomega and (0.01 -- $2) \times 10^{-6}$ for \omegaphi~\cite{VVBRAcp}.  Previous limits on these
branching fractions are presented in Refs.~\cite{OldCLEO,oldOmOmPhi}.  The results from Ref.~\cite{oldOmOmPhi},
$\Bomegaomega < 4.0 \times 10^{-6}$ and $\Bomegaphi < 1.2 \times 10^{-6}$, are based on about half the final 
\babar\ data sample.  In this Letter, we update the results of Ref.~\cite{oldOmOmPhi} using the final 
\babar\ dataset and improved analysis techniques.


Due to the limited size of the data sample, there is insufficient precision to determine the 
decay polarization in \omegaomega or \omegaphi.  We therefore integrate over the angular distributions,
%
correcting for
detector acceptance and efficiency.
The angular distribution is
\begin{eqnarray}
    &&\!\!\!\!\!\!\!\!\!\!\frac{1}{\Gamma}\frac{d^2\Gamma}{d\cos{\theta_{V_1}}d\cos{\theta_{V_2}}}
    =     \label{eq:vvAngDist} \\
&&\!\!\!\frac{9}{4}\left\{\frac{1}{4}(1-f_L)\sin^2{\theta_{V_1}}\sin^2{\theta_{V_2}}
     +f_L\cos^2{\theta_{V_1}}\cos^2{\theta_{V_2}}\right\}\! , \nonumber
\end{eqnarray}
where $V_{1,2} = \left(\{\omega,\omega\}\right.$ or $\left.\{\omega,\phi\}\right)$ are vector mesons, 
$\theta_{V_{1,2}}$ are helicity angles,
and $f_L$ is the fraction of events with longitudinal spin polarization.
For the $\phi$ meson, $\theta_\phi$ is the angle in the $\phi$ rest frame between the positively charged kaon
and the boost from the $B$ rest frame,
whereas for the $\omega$, $\theta_\omega$ is
the angle in the $\omega$ rest frame between the normal to the $\omega$ decay plane and the boost from the $B$ rest frame.
For both \omegaomega and \omegaphi, $f_L$ is predicted to be 80\% or larger~\cite{lilu}.


The data were collected
with the \babar\ detector~\cite{BABARNIM}
at the PEP-II asymmetric-energy $e^+e^-$ collider
located at the SLAC National Accelerator Laboratory.  An integrated
luminosity of 429~fb$^{-1}$~\cite{BabarLumi}, corresponding to 
$N_{\BB} = (471 \pm 3)\timesix$\ \BB\ pairs, was recorded at the $\Upsilon (4S)$
resonance (center-of-mass energy $\sqrt{s}=10.58\ \gev$).
Charged particles 
are detected, and their
momenta measured, by five layers of double-sided
silicon microstrip detectors 
and
a 40-layer drift chamber, both operating in the 1.5 T magnetic
field of a superconducting solenoid. We identify photons and electrons 
using a CsI(Tl) electromagnetic calorimeter.  
Charged particle identification (PID) is provided by energy loss measurements
in the tracking detector and by a ring-imaging Cherenkov detector.

We reconstruct the vector-meson decays through the \omtoppp\ and $\phi \ra \Kp\Km$
channels, with $\piz\ra\gaga$.  The minimum laboratory energy (momentum) required
for photons (charged kaons) is 50~MeV (100~MeV).  There is no specific minimum
momentum requirement for charged pions but they generally respect $p_T > 50$~MeV.
Charged pion and kaon candidates are rejected if their PID signature satisfies
tight consistency with protons or electrons, and the 
kaons must have a kaon signature, while the pions must not.  We require all
charged particle products associated with the $B$ meson candidate decay to
be consistent with having originated at a common vertex.

We apply the invariant mass requirements listed in Table~\ref{tab:rescuts}\ for the
\piz, $\omega$, and $\phi$ mesons.  After selection, the \piz is constrained to its
nominal mass~\cite{PDG2012}, which improves the $\omega$ mass resolution.  The
restrictions on the $\omega$ and $\phi$ meson masses are loose enough to
incorporate sideband regions.

\begin{table}[btp]
\begin{center}
\caption{
Selection requirements on the invariant mass of $B$-daughter intermediate states.  
}
\label{tab:rescuts}
\begin{tabular}{lc}
\dbline
State           & Inv.~mass (MeV)                 \\
\tbline                                        
\piz            & $120 < m_{\gamma\gamma} < 150$     \\
$\omega$        & $740 < m_{\pi\pi\pi} <820$         \\
$\phi$            & $1009<m_{KK} <1029$              \\
\dbline
\end{tabular}
\vspace{-5mm}
\end{center}
\end{table}

A $B$ meson candidate is characterized kinematically by the
energy-substituted mass $\mes=\sqrt{(\half
s+\pvec_0\cdot\pvec_B)^2/E_0^{*2}-\pvec_B^2}$ and the energy
difference $\DE = E_B^*-\half\sqrt{s}$, 
where $(E_0,\pvec_0)$ and $(E_B,\pvec_B)$ are the four-momenta of
the \UfourS\ and the $B$ candidate, respectively, 
and the asterisk denotes the \UfourS\ rest frame (quantities without asterisks are measured in the laboratory frame). 
For correctly reconstructed signal candidates, \DE\ and \mes\ peak at values of zero and $m_B$, 
respectively, with resolutions of about 30 MeV and 3.0 MeV.
Thus, signal events for this analysis mostly fall in the regions
$|\DE|\le0.1$ GeV and $5.27\le\mes\le5.29\ \gev$.
To incorporate sideband regions, we require
$|\DE|\le0.2$ GeV and $5.24\le\mes\le5.29\ \gev$.
The average number of candidates found per selected event is 
1.3 for \omegaphi decays and 1.7 for \omegaomega decays.  We 
choose the candidate with the smallest $\chi^2$ value constructed from the
deviations of the 
$\omega$ and $\phi$ resonance masses 
from their nominal values~\cite{PDG2012}.

Backgrounds arise primarily from random combinations of particles in
continuum events ($\epem\ra\qqbar,$ with $q=u,d,s,c$).  We reduce this background by
using the angle \thetaT\ in the \UfourS\ rest frame between the thrust axis~\cite{thrust} of the $B$
candidate and the thrust axis of the other charged and neutral particles in the event. 
The distribution of $|\costhr|$ is sharply peaked near $1.0$ for
\qqbar\ jet pairs, and nearly uniform for
$B$ meson decays.  
We require $|\costhr|<0.9$ for \omegaphi\ and
$|\costhr|<0.8$ for \omegaomega. 

We employ a maximum-likelihood fit, described below, to determine the
signal and background yields.  For the purposes of this fit, we construct
a Fisher discriminant~\cite{fisher} \xf\ that combines four variables defined
in the \UfourS\ frame: the polar angles with respect to the beam axis of the
$B$ meson momentum and $B$ thrust axis, and the zeroth and second angular
moments $L_{0}$ and $L_{2}$ of the energy flow about the $B$ thrust axis. 
The moments are defined by $ L_j = \sum_i
p_i\times\left|\cos\theta_i\right|^j,$ where $\theta_i$ is the angle
with respect to the $B$ thrust axis of a charged or neutral particle $i$,
$p_i$ is its momentum, and the sum excludes the $B$ candidate
daughters.

From simulated event samples produced with Monte Carlo (MC) event generators~\cite{geant},
we identify the most important backgrounds that arise from other \BB\ decay modes.  Most
of the \BB\ background does not peak in \mes\ or \DE\ and is grouped with continuum events
into a ``combinatoric'' background category.  Other \BB\ decay modes, such as $\Bz \to \omega\omega\piz$, 
$\Bz \to \omega\phi\piz$, $\Bz \to \omega\rho\pi$, $\Bz \to \omega a_1$, etc., peak in \mes\ and/or \DE\ 
and are referred to
as ``peaking'' background.  All peaking modes are grouped together into a single background component,
with a broad peak centered at negative values of \DE, and which is fitted in data simultaneously with the signal and 
combinatoric background components.

We obtain signal and background yields 
from extended unbinned 
maximum-likelihood fits with input observables \DE, \mes, \xf, and, for the vector meson 
$V = \omega$ or $\phi$, 
the mass $m_V$ and the cosine of the helicity angle $\cos \theta_V$.
For each $\omega$ meson, there is an additional helicity angle input observable, 
$\cos \Phi_{\omega}$, 
provided by the
polar angle, with respect to the $\omega$ flight direction, of the \piz in the 
\pip\pim
rest frame.  This angle is uncorrelated with the other
input observables
and has a distribution that is proportional to 
$\sin^2 \Phi_{\omega}$ 
for signal.  For background, the angular distribution is 
nearly flat in 
$\cos \Phi_{\omega}$,
and its deviation from flatness is parameterized
by separate third-order polynomials for combinatoric and for peaking \BB\ backgrounds.
For each event $i$ and component $j$ (signal, combinatoric background, 
peaking \BB\ background) we define the probability density function (PDF)
\begin{eqnarray}
\calP^i_j & = & \calP_j(\mes^i) \calP_j(\DE^i) \calP_j(\xf^i) 
\times \nonumber\\
  & & 
\calP_j(m^i_{V_1},m^i_{V_2},\cos\theta^i_{V_1},\cos\theta^i_{V_2}) \times \\
  & & 
\calP_j(\cos\Phi^{i}_{\omega_1})\calP_j(\cos\Phi^{i}_{\omega_2}), \nonumber
    \label{eq:evtL}
\end{eqnarray}
where the last of the $\calP_j$ terms is not present for \omegaphi.
The likelihood function is
\begin{equation}
    {\cal L} = \frac{e^{-(\sum Y_j)}}{N!} \prod_{i=1}^N
\sum_j Y_j {\cal P}_j^i\ , 
    \label{eq:totalL}
\end{equation}
where $Y_j$ is the event yield for component $j$
and $N$ is the number of events in the sample.  
  
\setlength{\tabcolsep}{2mm}
\begin{table*}[!bth]
\caption{
Fitted signal yield $Y_{\rm sig}$ and its statistical uncertainty, signal yield bias $Y_{\rm sig}^{\rm bias}$,
peaking \BB\ and combinatoric background yields $Y_{\rm peak}$ and $Y_{\rm comb}$ and their statistical uncertainties,
signal detection efficiency $\epsilon$ and its statistical uncertainty, daughter branching fraction product
$\prod\calB_i$ and its total uncertainty, significance $S$ (with systematic uncertainties included), 
measured branching fraction \calB\ (bold if evidence for signal is seen), and 90\% CL upper limit 
(UL, bold if no evidence) for 
the \omegaomega\ and \omegaphi\ decay modes.
}
\label{tab:results}
\begin{tabular}{lrrccrccccc}
\dbline
Mode            & \multicolumn{1}{c}{~~$Y_{\rm sig}$} &$Y_{\rm sig}^{\rm bias}$~ & $Y_{\rm peak}$ & $Y_{\rm comb}$ & \multicolumn{1}{c}{$\epsilon$} &$\prod\calB_i$ & $S$       &  \calB        & \calB\ UL \\
                & (events)                            & (events)         & (events)  & (events)     & \multicolumn{1}{c}{(\%)}       & (\%) &~~($\sigma$)~~ & $(10^{-6})$   & $(10^{-6})$ \\
\tbline
\fomegaomega   & $69.0^{+16.4}_{-15.2}$               & $7.3$~~~         & $3810 \pm 260$    & $53390 \pm 340$       & $14.0 \pm 0.1$ & $77.5 \pm 1.2$ & 4.4 & {\boldmath \romegaomega} & \ulomegaomega \\
\tbline
\fomegaphi     & $-2.8^{+5.7}_{-4.0}$                 & $-2.9$~~~        & $473^{+84}_{-80}$ & $17730^{+160}_{-150}$ &  $8.7 \pm 0.1$ & $43.2 \pm 0.6$ & 0.0 & \romegaphi  & {\boldmath \ulomegaphi} \\
\dbline
\end{tabular}
\vspace{-5mm}
\end{table*}
\setlength{\tabcolsep}{1mm}

For signal events, the PDF factor 
\begin{equation}
\calP_{\rm sig}(m_{V_1}^i,m_{V_2}^i,\cos\theta^i_{V_1},\cos\theta^i_{V_2})\nonumber
\end{equation}
takes the form
\begin{equation}
\calP_{1,{\rm sig}}(m_{V_1}^i)\calP_{2,{\rm sig}}(m_{V_2}^i){\cal Q}(\cos\theta^i_{V_1},\cos\theta^i_{V_2}),
\end{equation}
where ${\cal Q}$ 
corresponds to the right-hand side of Eq.~(\ref{eq:vvAngDist}) 
after modification to account for detector acceptance.  For 
combinatoric background events, the PDF factor is given for each vector meson independently by
\begin{eqnarray}
& & \!\!\!\!\!\!\!\!\!\!\!\! \calP_{\rm cont}(m_V^i,\cos\theta_V^i) = \nonumber\\
& & \!\!\!\!\!\!\!\!\!\! \calP_{\rm peak}(m_V^i)\calP_{\rm peak}(\cos\theta_V^i) + \calP_{\rm cont}(m_V^i)\calP_{\rm cont}(\cos\theta_V^i),\quad
\end{eqnarray}
distinguishing between genuine resonance ($\calP_{\rm peak}$) and
combinatorial ($\calP_{\rm cont}$) components.  
The background PDFs 
$\calP_{\rm peak}(\cos\theta_V^i)$ and $\calP_{\rm cont}(\cos\theta_V^i)$
are given by separately fitted third-order polynomials.
For the peaking \BB\ background, we assume that all four mass and helicity angle observables
are independent.

To describe the PDFs for signal, we use the sum of two Gaussians for  
${\cal P}_{\rm sig}(\mes)$ and for ${\cal P}_{\rm sig}(\DE)$.
An asymmetric Gaussian is used for ${\cal P}_{\rm sig}(\xf)$, i.e., two half-Gaussian distributions (one on the right
side of the mean and one on the left side) with different values for the standard deviation, summed with a small
additional Gaussian component to account for misreconstructed signal events.
The \mes, \DE, and \xf\ PDFs for peaking \BB background have the same functional form as for signal events, but their parameters 
are determined separately.
The genuine resonance components of ${\cal P}_j(m_V)$ are both described by 
relativistic Breit-Wigner distributions, each convolved with the sum of two
Gaussians to account for detector resolution, while the combinatoric components
of ${\cal P}_j(m_V)$ are described by third-order polynomials.  For the combinatoric
background category, the \mes\ distribution is described by
an ARGUS function $\mathcal{A}(\mes) \propto x\sqrt{1-x^2}\exp{\left[-\xi(1-x^2)\right]}$ (with $x\equiv\mes/E_B^*$)~\cite{ARGUS}, 
the \DE\ distribution by a second-order polynomial, and the \xf\ distribution by an
asymmetric Gaussian summed with an additional Gaussian.
The background PDF parameters that are
allowed to vary in the fit are the ARGUS function parameter $\xi$ for \mes, the
polynomial coefficients describing the combinatorial and the peaking \BB\ components for \DE\ and $m_V$, and the
\BB\ peak position and 
the two standard-deviation parameters of the asymmetric Gaussian for \xf.

For signal events, the PDF parameters are determined from simulation.
We study large control samples of $B \to D^{(*)}X$ events
with similar topology to the signal modes, such as
$\Bz \to \Dm\rhop$, to verify the simulated resolutions in \mes\ and \DE.
We make (small) adjustments to the signal PDFs to account for any
differences that are found.

In the fit to data, 13 parameters (out of around 130) are allowed to vary for each mode
including the yields $Y_j$ of the signal, total peaking \BB\ background, and total combinatoric 
background, and ten parameters of the continuum background PDFs.
For both modes, 
we set $f_L$ to 0.88, a value consistent with 
theoretical expectation~\cite{lilu}.
The event yields with their statistical uncertainties are presented in Table~\ref{tab:results}.

We evaluate possible biases in the signal yields, 
which might arise as a consequence of neglected correlations between the discriminating
variables, by applying our fit to an ensemble of simulated experiments.  
The numbers of signal and peaking \BB\ background events in these samples are Poisson-distributed
around the observed values and are extracted randomly from MC samples that include
simulation of the detector.
The largest of the correlations (approximately 15\%) is between the analysis variables
\mes and \DE.  
The signal yield bias $Y_{\rm sig}^{\rm bias}$ we find for each mode is provided in Table~\ref{tab:results}.  

The resulting branching fractions are calculated as 
\begin{equation}
{\cal B} = \frac{Y_{\rm sig} - Y_{\rm sig}^{\rm bias}}{\epsilon N_{\BB}},
\end{equation}
where the signal efficiencies $\epsilon$ are evaluated using MC and data control samples.  The total number of \BB\ pairs in
data $N_{\BB}$ is evaluated using a dedicated analysis~\cite{BCounting}.

The systematic uncertainties on the branching fractions 
are summarized in Table~\ref{tab:systematics}.
The uncertainty attributed to the yield-bias correction is taken to
be the quadrature sum of two terms: half of the bias correction and the
statistical uncertainty on the bias itself.  
The uncertainties of PDF parameters that are fixed in the fit are evaluated 
by taking the difference between the respective parameter values
determined in fits to simulated and observed $B \to D^{(*)}X$ events.  
Varying the signal PDF
parameters within these uncertainties, we estimate yield uncertainties for
each mode.  
Similarly, the uncertainty due to the modeling of the peaking \BB\ background
is estimated as the change in the signal yield when the number of peaking \BB\
background events is fixed (to within one standard deviation) of the expectation
from simulation.
We evaluate an uncertainty related to the constraint that all charged particles in the $B$
candidate emanate from a common vertex by the change in signal yield when this requirement
is removed.
The uncertainty associated with $f_L$ is evaluated by the change relative to
the standard result when $f_L$ is varied between the extreme values of 0.58
(the value of $f_L$ in $B \to \phi\Kstar$ decays) and 1.0.

\begin{table}[btp]
\begin{center}
\caption{
Estimated systematic uncertainties on the branching fractions \Bomegaomega\ and \Bomegaphi.
Additive and multiplicative uncertainties are independent and are combined in quadrature.
Note that only the additive uncertainties are consequential in the case of the \omegaphi mode, as essentially
zero signal is observed in that mode.
}
\label{tab:systematics}
\begin{tabular}{lcc}
\dbline
Decay Mode           & \omegaomega     &    \omegaphi    \\
\tbline                                        
  Additive uncertainties (events):     &          &             \\
~~Fit bias                      &   5.5    &     2.0     \\
~~Fit parameters                &   0.5    &     0.3     \\
~~\BB backgrounds               & $\!\!\!\!\!\!< 0.1$  &  $\!\!\!\!\!\!< 0.1$    \\
  Total additive (events)       &   5.5    &     2.0     \\
\tbline
  Multiplicative uncertainties (\%):   &             &             \\
~~$f_L$ variation               & +25.3 -8.3  & +18.3 -48.0 \\
~~Vertex finding efficiency     & +5.3 -0.0   & +25.0 -50.0 \\
~~Track finding efficiency      &   1.0    &      1.0    \\
~~\piz efficiency               &   4.2    &      2.1    \\
~~Kaon identification           &   ---    &      4.5    \\
~~\costhr\ cut efficiency       &   1.3    &      1.4    \\
~~Submode branching fractions   &   1.6    &      1.5    \\
~~MC statistics                 &   1.0    &      1.4    \\
~~Total number of \BB in data   &   0.6    &      0.6    \\
  Total multiplicative (\%)     & +26.3 -9.7  & +31.5 -69.5 \\
\dbline
\end{tabular}
\vspace{-5mm}
\end{center}
\end{table}

Systematic uncertainties associated with the selection efficiency, evaluated
with data control samples,
are 
$0.8\%\times N_t$ and $3.0\%\times N_{\pi^0}$, where $N_t$ is the number of
tracks and $N_{\pi^0}$ the number of $\pi^0$ mesons~\cite{TrackEff}.
The uncertainty of $N_{\BB}$ is 0.6\%~\cite{BCounting}.  
World averages~\cite{PDG2012}\ provide the uncertainties in the $B$-daughter product
branching fractions (1--2\%).  
The uncertainty associated with the requirement on \costhr\ is 1--2\% depending on the decay mode.

\begin{figure}[!htbp]
\begin{center}
  \includegraphics[width=0.49\linewidth]{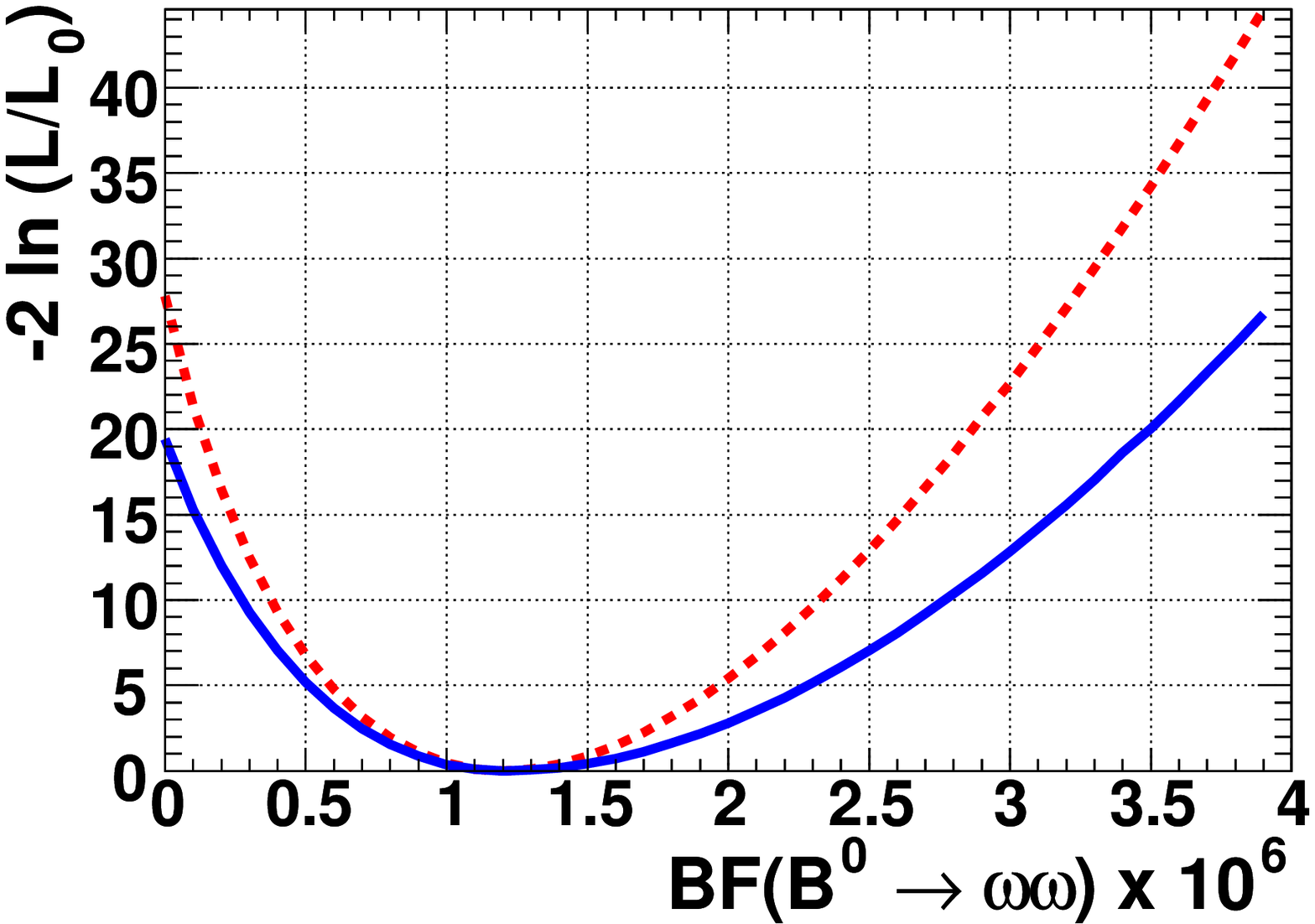}
\hfill
  \includegraphics[width=0.49\linewidth]{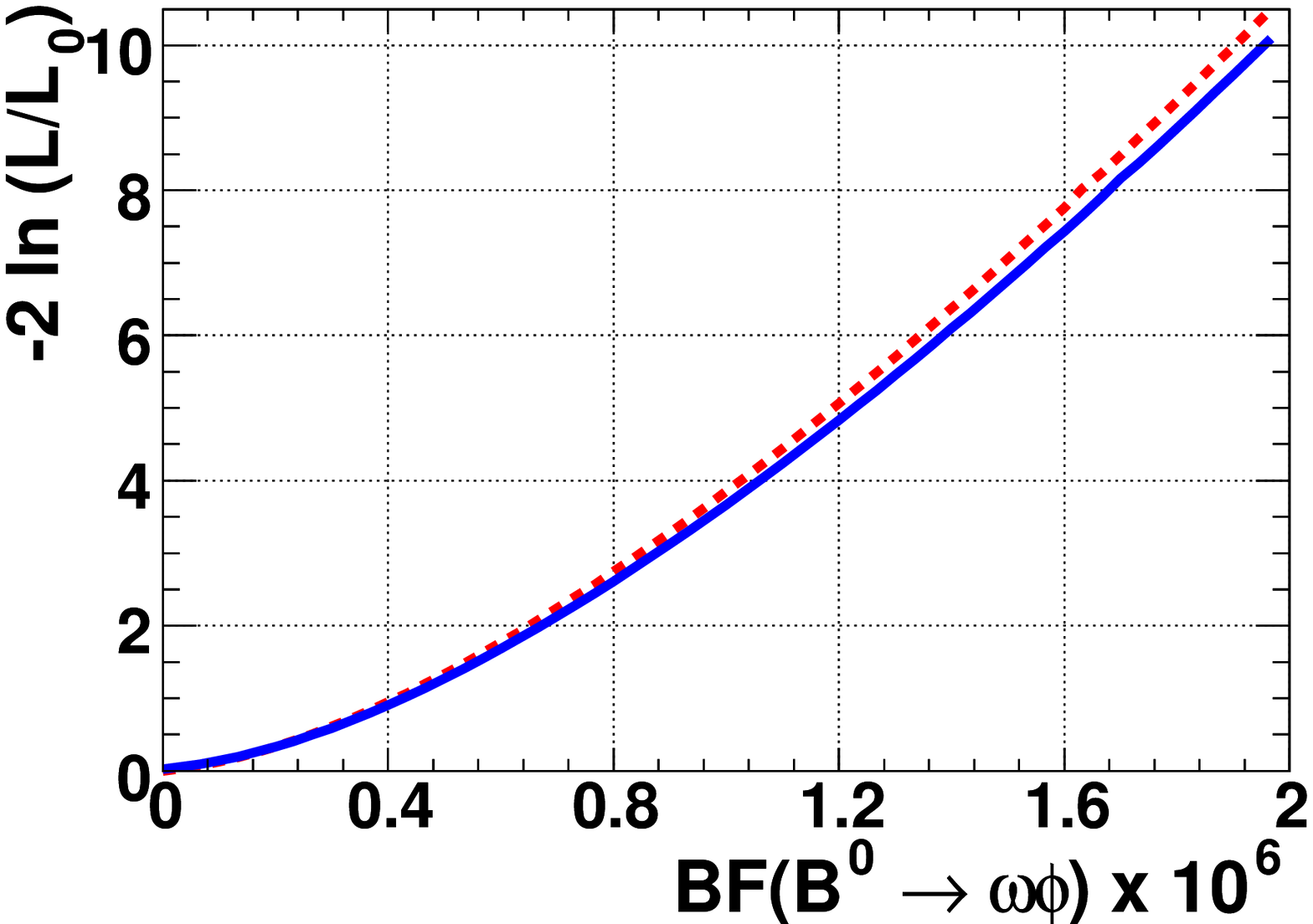}
  \caption{Distribution of $-2\ln{\calL}(\calB)$ (normalized to the maximum likelihood $\calL_0$) for \omegaomega (left)
  and \omegaphi (right) decays.  The dashed curves include only statistical uncertainties;
  the solid curves include systematic uncertainties as well.}  
  \label{fig:BFscans}
\end{center}
\end{figure}

Table~\ref{tab:results} also presents the measured branching fractions,
total associated uncertainties, and significances.
The significance, which we denote in terms of the analogous number of Gaussian standard deviations, 
is taken as the square root of the difference between the
value of $-2\ln{\cal L}$ (with systematic uncertainties included) for
zero signal events and the value at its minimum.  
The behavior of $-2\ln{\calL}(\calB)$ for the two modes
is shown in Fig.~\ref{fig:BFscans}.  We find evidence for 
\omegaomega\ decays at the level of 4.4 standard deviations including systematic uncertainties.
For each mode
we also quote a 90\%\ CL upper limit,
taken to be the branching fraction below which lies 90\% of the total
of the likelihood integral in the positive branching fraction region.
In calculating branching fractions we assume that the decay rates of
the \UfourS\ to \BpBm\ and \BzBzb\ are equal \cite{PDG2012}.  

\begin{figure}[!htbp]
\hspace*{-1.2mm}
\rotatebox{90}{Events / 2.5 MeV}
\hspace*{-1.3mm}
  \includegraphics[trim = 0mm 0mm 0mm 0mm, clip, width=0.42\linewidth]{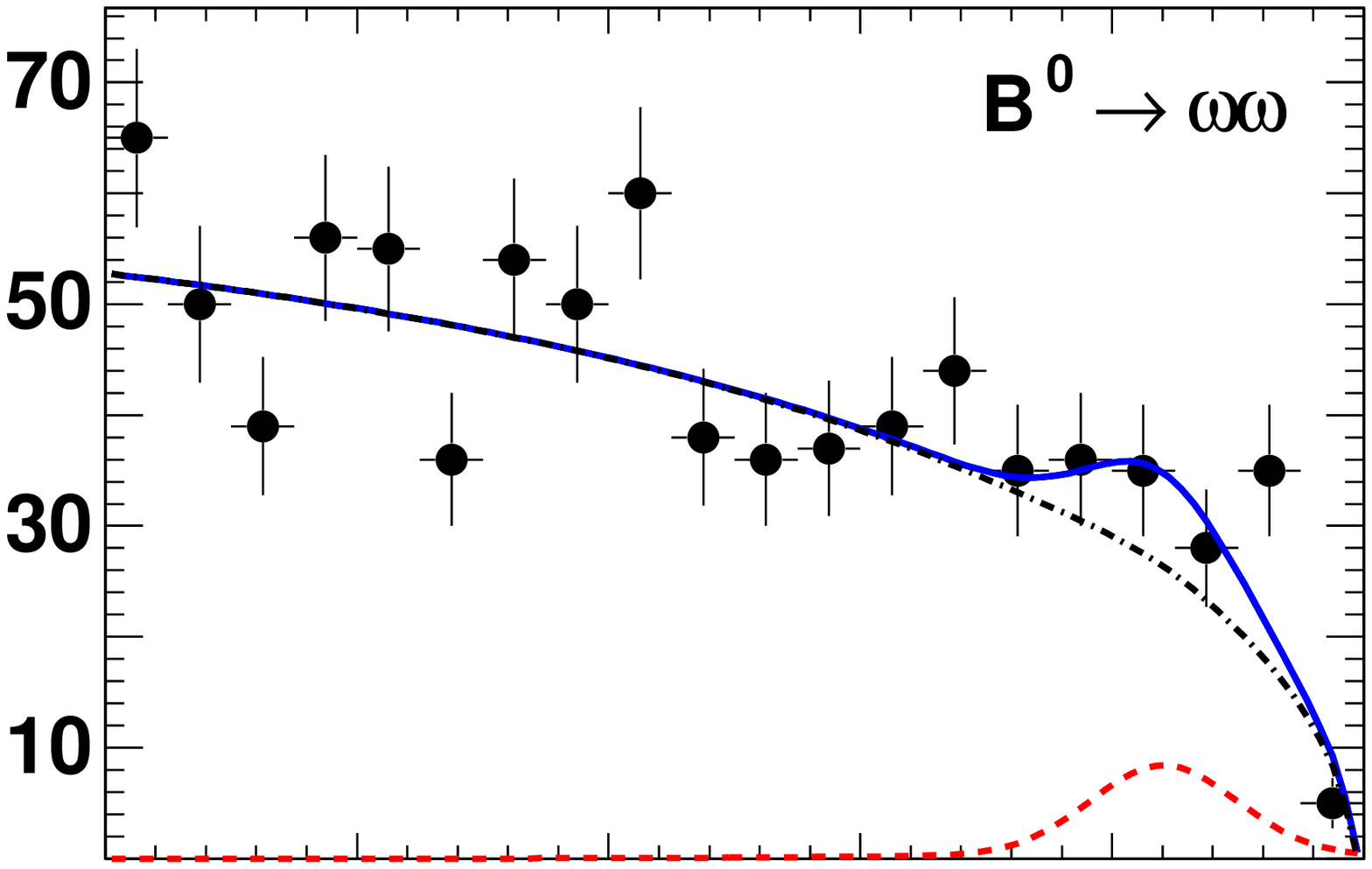}
\hspace*{2.3mm}
\rotatebox{90}{Events / 20 MeV}
\hspace*{-2.2mm}
  \includegraphics[trim = 0mm 0mm 0mm 0mm, clip, width=0.424\linewidth]{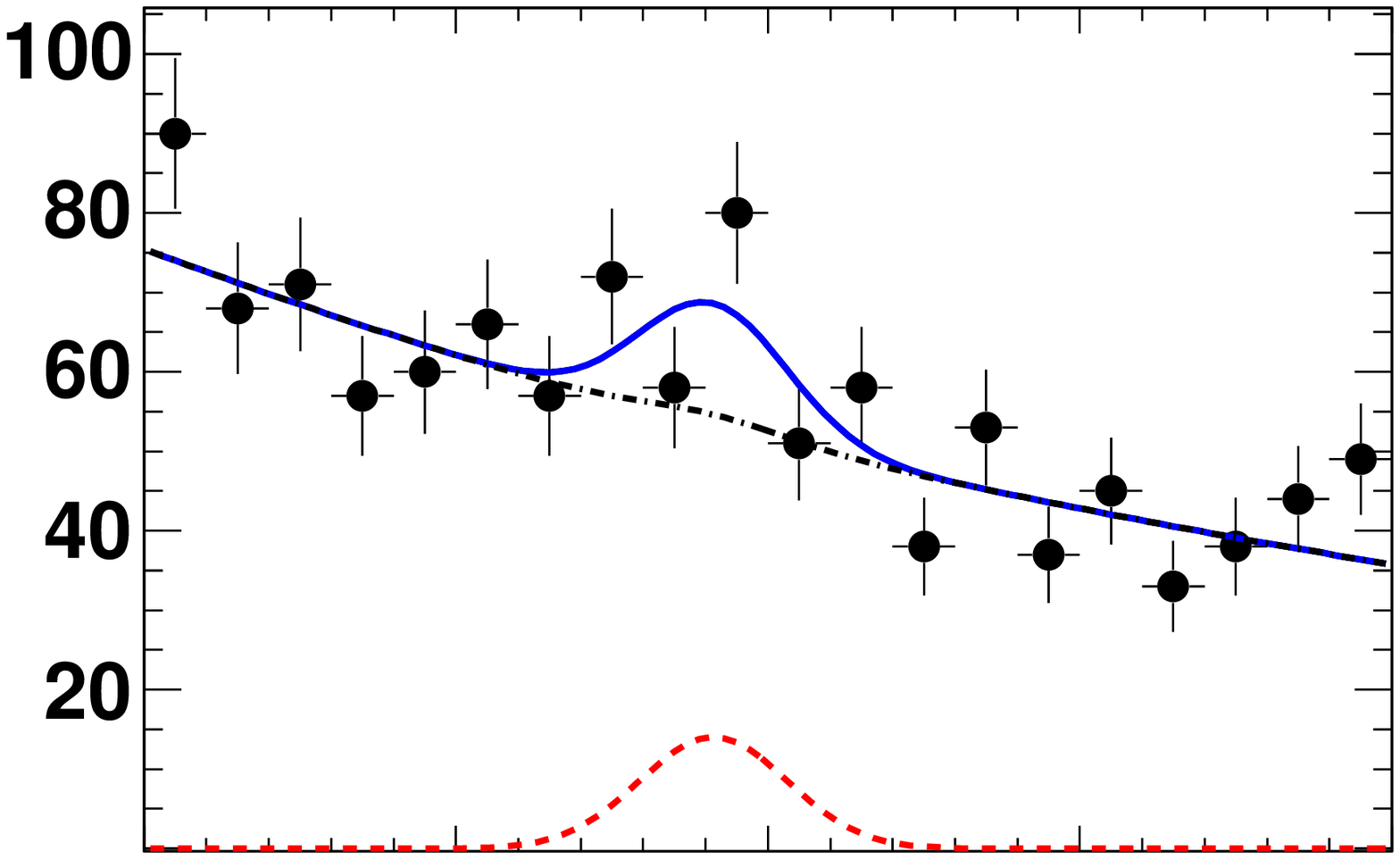}
\hfill

\vspace*{-0.4mm}

\hspace*{3.35mm}
  \includegraphics[trim = 0mm 0mm 0mm 0mm, clip, width=0.442\linewidth]{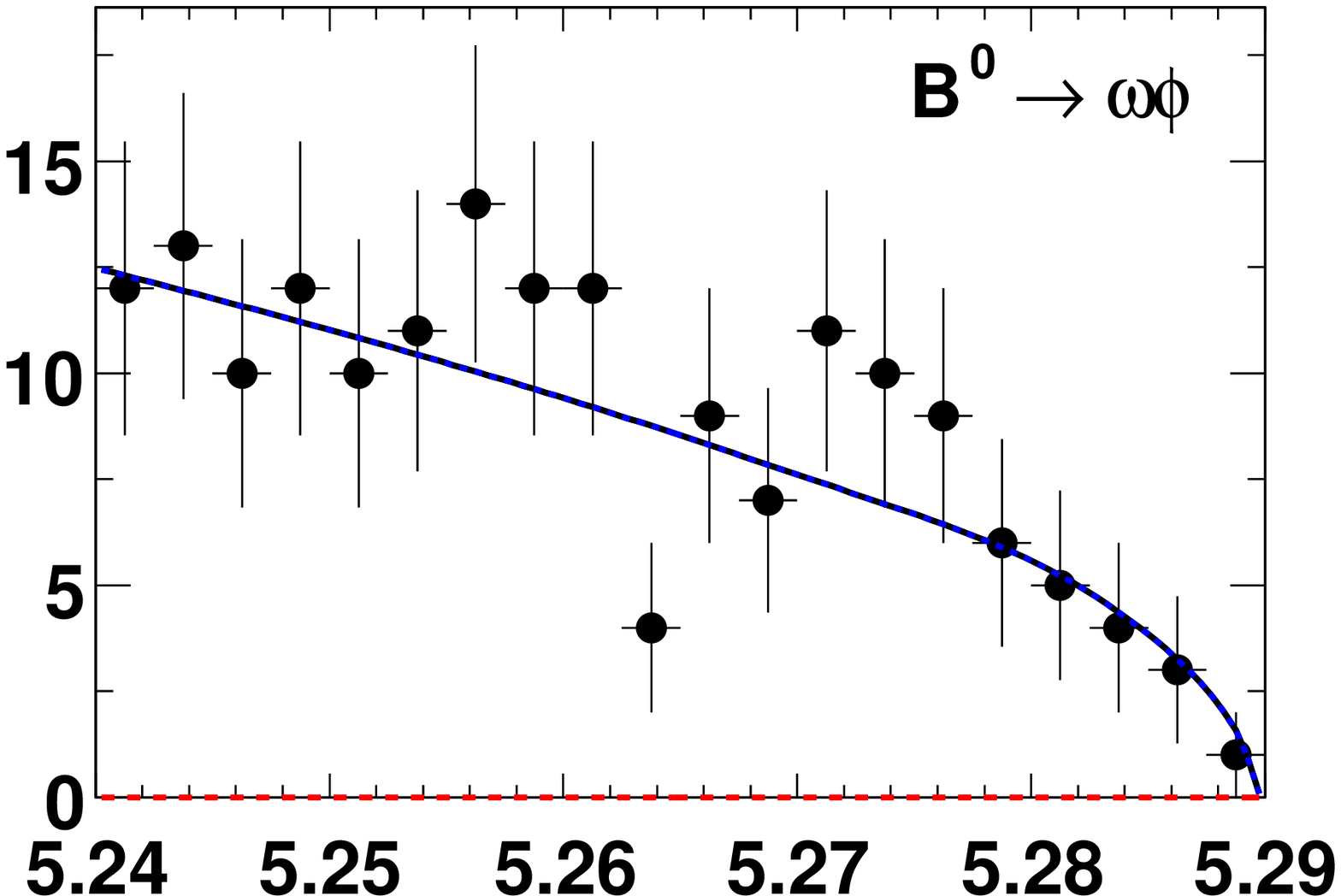}
\hfill
  \includegraphics[trim = 0mm 0mm 0mm 0mm, clip, width=0.429\linewidth, height=25.6mm]{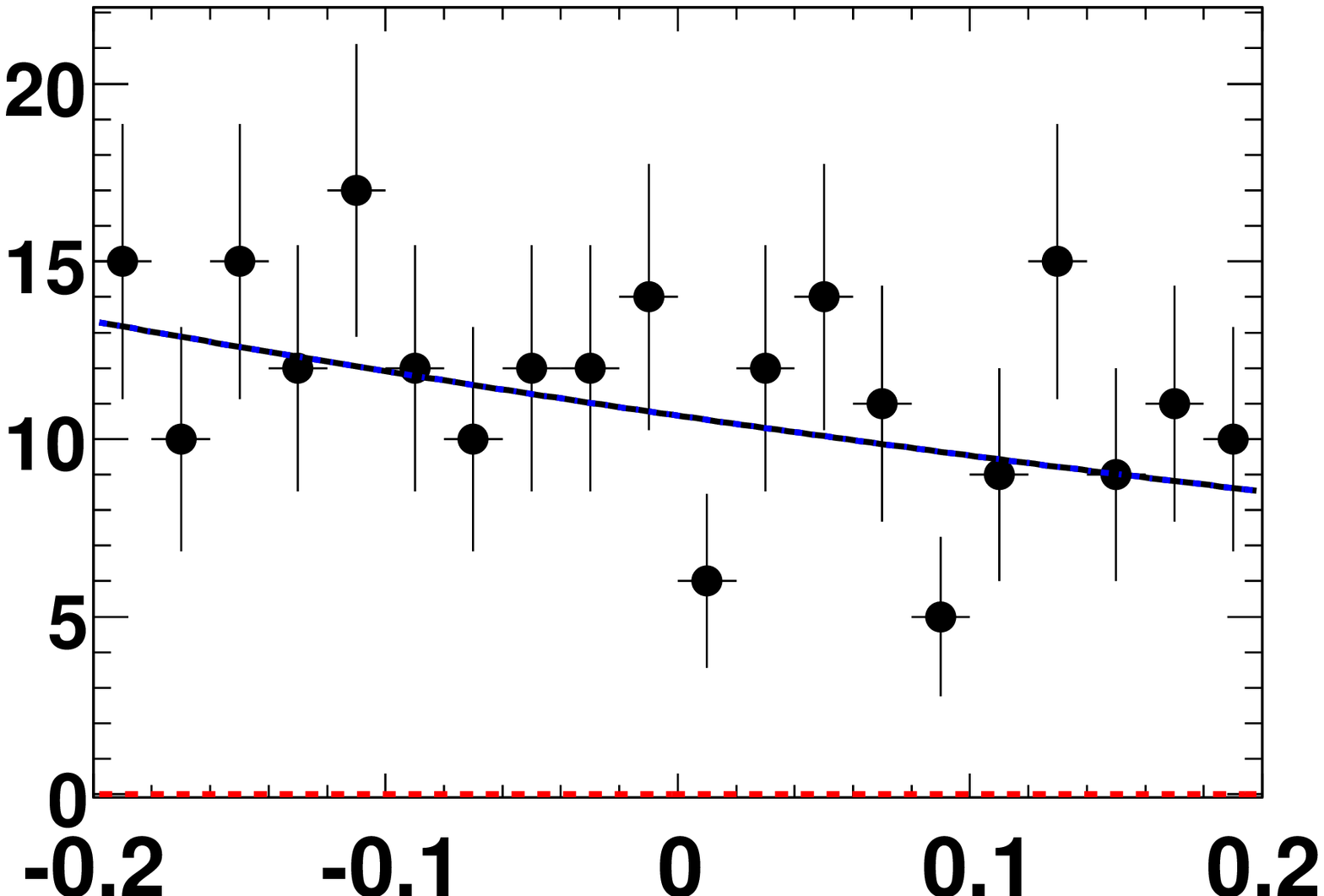}
\hspace*{2.1cm} \textbf{\boldmath \mes}(GeV) \hspace*{2.7cm} \textbf{\boldmath \DE}~(GeV)

  \caption{Projections of  \mes\ (left) and \DE\ 
    (right) for 
a signal-enriched sample of
events passing a 
    set of dedicated selection cuts 
    for \omegaomega (upper plots), and \omegaphi (lower plots).  
    The solid curve gives the total PDF (computed
    without the variable plotted), the dashed curve is the signal
    contribution, and the dot-dashed curve is the background
    contribution, which includes both combinatoric and peaking \BB\ backgrounds.  
}
  \label{fig:proj_omegaKst}
\end{figure}

Figure~\ref{fig:proj_omegaKst} presents
the data and PDFs projected onto \mes\ and \DE, for subsamples enriched 
with signal events
via a set of selection criteria on the analysis variables.
The selection criteria 
are
$|m_{\omega} - m_{\omega}^{\rm nominal}| <$ 15~MeV, $|m_{\phi} - m_{\phi}^{\rm nominal}| <$ 8~MeV,
$\mathcal{F} <$ 0.1, 
$|\cos\Phi_{\omega}| <$ 0.95, and
$|\costhr| <$ 0.8, with 
$|\DE| <$ 30~MeV for the two \mes\ plots and 
$\mes >$ 5.274~GeV for the two \DE\ plots.
These criteria retain 23\% (40\%) of 
\omegaomega\ (\omegaphi) signal events, and
in both modes reject over 99\% of the background events.

In summary, 
we have performed searches for 
\omegaomega\ and $\omega\phi$ decays.
We establish the following branching fraction and upper limit: 
\begin{eqnarray}
    \Bomegaomega&\! = \! & (\romegaomega) \times10^{-6}\;\;\;\mbox{and}\nonumber\\
    \Bomegaphi  &\! < \! & ~~~\ulomegaphi \times10^{-6}~\mbox{(90\% CL)}.\nonumber
\end{eqnarray}
For the branching fraction, the first uncertainty is statistical and the second is systematic.
These results provide the first evidence for \omegaomega\ decays and 
improve the constraint on the \omegaphi\ branching fraction.
Our results are in agreement with
theoretical estimates \cite{VVBRAcp,lilu}.

We are grateful for the excellent luminosity and machine conditions
provided by our \pep2\ colleagues, 
and for the substantial dedicated effort from
the computing organizations that support \babar.
The collaborating institutions wish to thank 
SLAC for its support and kind hospitality. 
This work is supported by
DOE
and NSF (USA),
NSERC (Canada),
CEA and
CNRS-IN2P3
(France),
BMBF and DFG
(Germany),
INFN (Italy),
FOM (The Netherlands),
NFR (Norway),
MES (Russia),
MINECO (Spain),
STFC (United Kingdom),
BSF (USA-Israel). 
Individuals have received support from the
Marie Curie EIF (European Union)
and the A.~P.~Sloan Foundation (USA).


\end{document}